\documentstyle[12pt]{article}
\parskip=\medskipamount                 % space between paragraphs
\thicklines                     % thick straight lines for pictures
\newcommand{\bim}[6]{\bibitem{#1}#2, {\em #3\/}$\;${\bf
#4}$\;$(#5)$\;${#6}.}
\def\ZZ{\relax{\sf Z\kern-.4em Z}}
\def\a{\alpha} \def\b{\beta}   \def\e{\epsilon} \def\g{\gamma}
 \def\l{\lambda} 
   
   \def\cD{{\cal D}}
   
 \def\cK{{\cal K}} \def\cL{{\cal L}} \def\cM{{\cal M}}
 \def\cO{{\cal O}}  

\def\etc{{\it etc\/}}
\newtheorem{proposition}{Proposition}[section]
\newtheorem{corollary}{Corollary}[section]

\newlength{\shiftwidth}
\def\shift#1{&&\hbox to \shiftwidth{\hfill $\displaystyle#1$}}
\newlength{\sshiftwidth}
\def\sshift#1{\lefteqn{\hbox to
\sshiftwidth{\hfill$\displaystyle#1$}}}

\def\ba{\mbox{\boldmath$\alpha$}}
\def\bb{\mbox{\boldmath$\beta$}}
\def\bl{\mbox{\boldmath$\lambda$}}
\def\br{\mbox{\boldmath$\rho$}}
\def\bn{\mbox{\boldmath$n$}}

\def\sba{\mbox{\boldmath$\alpha$}}
\def\sbb{\mbox{\boldmath$\beta$}}

\def\sign#1{{\rm sign}\left(#1\right)}

\def\rank{{\rm rank}\,}
\def\dim{{\rm dim}\,}
\def\Vol{{\rm Vol}\,}
\def\ord{{\rm ord}\,}

\def\Tr{{\rm Tr}}
\def\Pexp{{\rm Pexp}}
\def\hol#1{\Pexp\left(\oint_{#1} A_\mu dx^\mu\right)}
\def\thol#1#2{\Tr_{#1}\hol{#2}}

\def\TK{{\rm Tub}(\cK)}
\def\Upq{U^{(p,q)}}
\def\tu{\tilde{U}^{(p,q)}_{\a 1}}
\def\PU{\Phi(\Upq)}
\def\PUm{\Phi(U^{(-q_j,p_j)})}

\def\eba{\exp\left(\frac{2\pi i}{K}\ba\right)}

\def\east{\exp\left(\frac{2\pi i}{K}\a_*\right)}

\def\pf{\phi_{\rm fr}}
\def\Nph{N_{\rm ph}}
\def\xgf{X_g\left(\frac{p_1}{q_1},\ldots,\frac{p_n}{q_n}\right)}

\def\pqn{\left(\frac{p}{q}+\nu\right)}
\def\spqn{\sign{\frac{p}{q}+\nu}}
\def\tspqn{3\sign{\frac{p}{q}+\nu}}

\def\zmk#1{Z(M^{#1};k)}
\def\zamk{Z_{\a} (M,\cK;k)}
\def\zbamk{Z_{\ba} (M,\cK;k)}
\def\zas{Z_\a (S^3,\cK;k)}
\def\ztramk{Z^{({\rm tr})}_{\a}(M,\cK;k)}
\def\ztrbamk{Z^{({\rm tr})}_{\a} (M,\cK;k)}
\def\ztrmk{Z^{({\rm tr})}(M;k)}
\def\ztrmpk{Z^{({\rm tr})}(M^\prime;k)}

\def\dask{\Delta_A(S^3,\cK;\exp(2\pi ia))}
\def\damk{\Delta_A(M,\cK;\exp(2\pi ia))}
\def\trma{\tau_R(M\setminus\TK;\exp(2\pi ia))}
\def\trsa{\tau_R(S^3\setminus\TK;\exp(2\pi ia))}

\def\rhs{RHS$\;$}
\def\rhss{RHS}

\def\ohm{\ord H_1(M,\ZZ)}
\def\eamd{\exp\left(2\pi i\frac{a}{m_2 d}\right)}
\def\ep{e^{-\pi\epsilon\a^2}}
\def\lime{\lim_{\epsilon\rightarrow 0}}

\def\Res{{\rm Res}}

\catcode`\@=11

\newif\if@fewtab\@fewtabtrue

\catcode`\@=11

\newif\if@fewtab\@fewtabtrue

{\count255=\time\divide\count255 by 60
\xdef\hourmin{\number\count255}
\multiply\count255 by-60\advance\count255 by\time
\xdef\hourmin{\hourmin:\ifnum\count255<10 0\fi\the\count255}}
\def\ps@draft{\let\@mkboth\@gobbletwo
    \def\@oddhead{}
    \def\@oddfoot
       {\hbox to 7 cm{$\scriptstyle Draft\ version:\ \draftdate$
       \hfil}\hskip -7cm\hfil\rm\thepage \hfil}
    \def\@evenhead{}\let\@evenfoot\@oddfoot}

\def\ceqno{\global\@fewtabfalse
    \ifcase\@eqcnt \def\@tempa{& & &}\or \def\@tempa{& &}
      \or \def\@tempa{&}
      \or\def\@tempa{}\fi\@tempa
{\rm(\theequation)}}

\def\aeqno#1{\global\@fewtabfalse
    \ifcase\@eqcnt \def\@tempa{& & &}\or \def\@tempa{& &}
      \or \def\@tempa{&}
      \or\def\@tempa{}\fi\@tempa
{\rm(\theequation,#1)}}

\def\label#1{\ifnum\draftcontrol=1
 \global\def\draftnote{$\scriptstyle #1$}\fi
 \@bsphack\if@filesw {\let\thepage\relax
   \def\protect{\noexpand\noexpand\noexpand}%
\xdef\@gtempa{\write\@auxout{\string
      \newlabel{#1}{{\@currentlabel}{\thepage}}}}}\@gtempa
   \if@nobreak \ifvmode\nobreak\fi\fi\fi
  \@esphack}

\def\alabel#1#2{\label{#1}\global\@fewtabfalse
    \ifcase\@eqcnt \def\@tempa{& & &}\or \def\@tempa{& &}
      \or \def\@tempa{&}
      \or\def\@tempa{}\fi\@tempa
{\hbox to 3cm{\phantom{\rm(\theequation,#2)}
\draftnote \hfil}\hskip -3cm {\rm(\theequation,#2)}}}

\def\clabel#1{\label{#1}\global\@fewtabfalse
    \ifcase\@eqcnt \def\@tempa{& & &}\or \def\@tempa{& &}
      \or \def\@tempa{&}
      \or\def\@tempa{}\fi\@tempa
{\hbox to 3cm{\phantom{\rm(\theequation)}
\draftnote \hfil}\hskip -3cm{\rm(\theequation)}}}

\def\eqnarray{\def\draftnote{{}}\global\@fewtabtrue
\stepcounter{equation}\let\@currentlabel=\theequation
\global\@eqnswtrue
\global\@eqcnt\z@\tabskip\@centering\let\\=\@eqncr
$$\halign to \displaywidth\bgroup\@eqnsel\hskip\@centering\@eqcnt\z@
  $\displaystyle\tabskip\z@{##}$&\global\@eqcnt\@ne
  \hskip 1\arraycolsep \hfil${##}$\hfil
  &\global\@eqcnt\tw@ \hskip 1\arraycolsep
$\displaystyle\tabskip\z@{##}$
\hfil  \tabskip\@centering&\global\@eqcnt\thr@@\llap{##}\tabskip\z@
\cr}

\def\endeqnarray{\@@eqncr\egroup
      \global\advance\c@equation\m@ne$$\global\@ignoretrue}

\def\@eqnnum{\hbox to 3cm{\phantom{\rm(\theequation)} \draftnote
                         \hfil}\hskip -3cm {\rm(\theequation)}}

\def\@@eqncr{\let\@tempa\relax
    \ifcase\@eqcnt \def\@tempa{& & &}\or \def\@tempa{& &}
      \or \def\@tempa{&}
      \or\def\@tempa{}
\fi\@tempa
\if@eqnsw
\if@fewtab\@eqnnum\fi
\stepcounter{equation}\fi\global
\@eqnswtrue\global\@eqcnt\z@\global\@fewtabtrue\cr}

\def\draftcite#1{\ifnum\draftcontrol=1#1\else{}\fi}

\def\@lbibitem[#1]#2{\item{}\hskip -3cm \hbox to 2cm
{\hfil$\scriptstyle\draftcite{#2}$}\hskip
1cm[\@biblabel{#1}]\if@filesw
     {\def\protect##1{\string ##1\space}\immediate
      \write\@auxout{\string\bibcite{#2}{#1}}}\fi\ignorespaces}

\def\@bibitem#1{\item\hskip -3cm \hbox to 2cm
{\hfil $\scriptstyle\draftcite{#1}$}\hskip 1cm
\if@filesw \immediate\write\@auxout
       {\string\bibcite{#1}{\the\value{\@listctr}}}\fi\ignorespaces}

\def\nsection#1{\section{#1}\setcounter{equation}{0}}
\def\nappendix#1{\def\thesection{A#1}\section*{Appendix #1}
\def\theequation{{A#1.\arabic{equation}}}
\def\theproposition{{A#1.\arabic{proposition}}}
\setcounter{equation}{0}
\setcounter{proposition}{0}}

\def\draftdate{\number\month/\number\day/\number\year\ \ \ \hourmin }

\global\def\draftcontrol{0}
\catcode`\@=12

\def\theequation{{\thesection.\arabic{equation}}}

\def\qq{\begin{eqnarray}}
\def\qqq{\end{eqnarray}}

\begin{document}
%\draft

\begin{titlepage}
\centerline{\hfill                 UMTG--172}
\centerline{\hfill                 UTTG--30--93}
\centerline{\hfill                 hep-th/9401061}
\vfill
\begin{center}
{\large \bf A Contribution of the Trivial Connection to Jones
Polynomial and Witten's Invariant of 3d Manifolds. I
} \\

\bigskip
\centerline{L. Rozansky\footnote{Work supported
in part by the National Science Foundation
under Grants No. PHY-92 09978 and 9009850 and by the R. A. Welch
Foundation.
}}

\centerline{\em Physics Department, University of Miami
}
\centerline{\em P. O. Box 248046, Coral Gables, FL 33124, U.S.A.}

\vfill
{\bf Abstract}

\end{center}
\begin{quotation}

We use a path integral formulation of the Chern-Simons quantum field
theory in order to give a simple ``semi-rigorous'' proof of a recently
conjectured limitation on the $1/K$ expansion of the Jones polynomial
of a knot and its relation to the Alexander polynomial.  A combination
of this limitation with the finite version of the Poisson resummation
allows us to derive a surgery formula for the contribution of the
trivial connection to Witten's invariant of rational homology spheres.
The 2-loop part of this formula coincides with Walker's surgery
formula for Casson-Walker invariant.  This proves a conjecture that
Casson-Walker invariant is proportional to the 2-loop correction to
the trivial connection contribution. A contribution of the trivial
connection to Witten's invariant of a manifold with nontrivial
rational homology is calculated for the case of Seifert manifolds.

\end{quotation}
\vfill
\end{titlepage}

\pagebreak
%\tableofcontents
%\pagebreak
%+++++++++++++++++++++++++++++++++++++++++++++++++++
\nsection{Introduction}
%+++++++++++++++++++++++++++++++++++++++++++++++++++
In his paper \cite{Wi1}, Witten defined a topological invariant of a
3d manifold $M$ with an $n$-component link $\cL$ inside it as a
partition function of a quantum Chern-Simons theory. Let us attach
representations $V_{\ba_i},\;1\leq i\leq n$ of a simple Lie group $G$
to the components of $\cL$ (in our notations $\ba_i$ are the highest
weights shifted by $\br=\frac{1}{2}\sum_{\bl_i\in\Delta_+}\bl_i$,
$\Delta_+$ is a set of positive roots of $G$). Then Witten's invariant
is equal to the path integral over all gauge equivalence classes of
$G$ connection on $M$:
\qq
Z_{\ba_1,\ldots,\ba_n}(M,\cL;k)=\int[\cD A_\mu]
\exp\left(\frac{i}{\hbar}S_{CS}\right)
\prod_{i=1}^{n}\thol{\ba_i}{L_i},
\label{1.1}
\qqq
here $A_\mu$ is a connection, $S_{CS}$ is its Chern-Simons action
\qq
S_{CS}=\frac{1}{2}\,\Tr\,\epsilon^{\mu\nu\rho}\int_{M}dx
(A_\mu \partial_\nu A_\rho + \frac{2}{3}A_\mu A_\nu A_\rho),
\label{1.2}
\qqq
$\Tr$ is a trace in the fundamental representation (so that $\Tr
\bl_i^2=2$ for long roots of $G$), $\hbar$ is a Planck's constant:
\qq
\hbar=\frac{2\pi}{k},\;\;k\in \ZZ.
\label{1.3}
\qqq
$\thol{\ba_i}{L_i}$ are the traces of holonomies along the link
components $L_i$ taken in the representations $V_{\ba_i}$. Witten
showed that for a link in $S^3$ his invariant was proportional to the
Jones polynomial of that link. In what follows we will refer to
eq.~(\ref{1.1}) as the definition of the Jones polynomial and its
normalization.

Witten derived a surgery algorithm for an exact calculation of the
path integral~(\ref{1.1}). We review it briefly in order to set our
notations. Consider a manifold $M$ with a knot $\cK$ inside it. Let us
choose a basis of cycles on the boundary of its tubular neighborhood
$\TK$. $C_1$ is a cycle contractible through the tubular neighborhood
(i.e. $C_1$ is the meridian of $\cK$). $C_2$ is a cycle which has a
unit intersection number with $C_1$ ($C_2$ is defined only modulo
$C_1$). Cut out the tubular neighborhood $\TK$ and glue it back in
such a way that the cycles $pC_1 + qC_2$ and $rC_1 + sC_2$ on the
boundary of the complement of $\cK$ are identified with the cycles
$C_1$ and $C_2$ on the boundary of $\TK$. As a result of this surgery,
a new manifold $M^\prime$ is constructed.

The integer numbers $p,q,r,s$ form a unimodular matrix
\qq
\Upq=\left(
\begin{array}{cc}
p&r\\q&s
\end{array}\right)\in SL(2,\ZZ),\;\;
ps-qr=1.
\label{1.4}
\qqq
The group $SL(2,\ZZ)$ has a unitary representation in the space of
affine characters of $G$ which is in fact a Hilbert space of the
Chern-Simons theory corresponding to $T^2=\partial\TK$. The basis
vectors of this space $|\ba,1\rangle$ ($\ba\in\Delta_G =\Lambda^w
/(W\times K\Lambda^{R})\setminus
{\rm walls}$, $K=k+c_V$, $c_V$ is a dual
Coxeter number of $G$, $c_V=N$ for $SU(N)$) are the eigenstates of the
holonomy operator along the cycle $C_1$:
\qq
\Pexp\left(\oint_{C_1}\hat{A}^\mu dx^\mu \right)
|\ba,1\rangle=
%\exp\left(\frac{2\pi i}{K}\ba\right)
\eba|\ba,1\rangle,
\label{1.5}
\qqq
here $\hat{A}^\mu$ is an operator corresponding to the classical field
$A^\mu$. The matrix elements of $\Upq$ represented in this basis are
(for a simply laced group)
\begin{eqnarray}
\tilde{U}^{(p,q)}_{\sba\sbb}&=&
\frac{[i\,\sign{q}]^{|\Delta_+|}}{(K|q|)^{\rank G/2}}
\exp\left[-\frac{i\pi}{12}\dim G \,\PU\right]
\left(\frac{\Vol \Lambda^w}{\Vol \Lambda^R}\right)
^{\frac{1}{2}}
%\nonumber
\label{1.6}
\\
&&\times\sum_{\bn\in \Lambda^R/q\Lambda^R}
\sum_{w\in W}(-1)^{|w|}\exp\frac{i\pi}{Kq}
\left[p\ba^2-2\ba\cdot(K\bn+w(\bb))+s(K\bn+w(\bb))^2\right],
\nonumber
%\label{1.6}
\end{eqnarray}
here $|\Delta_+|$ is a number of positive roots in $G$, $W$ is the
Weyl group and $\PU$ is the Rademacher function defined as follows:
\qq
\Phi\left[
\begin{array}{cc}p&r\\q&s\end{array}
\right]=\frac{p+s}{q}-12s(s,q),
\label{1.7}
\qqq
$s(s,q)$ is a Dedekind sum:
\qq
s(m,n)=\frac{1}{4n}\sum_{j=1}^{n-1}
\cot\left(\frac{\pi j}{n}\right)
\cot\left(\frac{\pi mj}{n}\right).
\label{1.8}
\qqq
The formula~(\ref{1.6}) was derived by L. Jeffrey \cite{Je} for
$G=SU(2)$:
\begin{eqnarray}
%\lefteqn{
&{\displaystyle
\tilde{U}^{(p,q)}_{\a\b}=i\frac{\sign{q}}{\sqrt{2K|q|}}
e^{-\frac{i\pi}{4}\PU}
%}
%\label{1.9}
%\\
%&&\times
\sum_{\mu=\pm 1}\sum_{n=0}^{q-1}
\mu\exp\frac{i\pi}{2Kq}\left[p\a^2-2\a(2Kn+\mu\b)
+s(2Kn+\mu\b)^2\right],
%}
}
&
\nonumber\\
&\Delta_{SU(2)}:\;\;1\leq\a,\b\leq K-1.&
\label{1.9}
%\nonumber
\end{eqnarray}

According to Witten \cite{Wi1}, the invariant of the manifold
$M^\prime$ constructed by a $\Upq$ surgery on a knot $\cK$ in a
manifold $M$ can be expressed through the Jones polynomial of that
knot and the representation~(\ref{1.8}) of the surgery matrix:
\qq
\zmk{\prime}=e^{i\pf}\sum_{\ba\in\Delta_G}
\zbamk\tilde{U}_{\ba\br}^{(p,q)}
\label{1.10}
\qqq
(recall that $\br$ is a shifted highest weight of the trivial
representation). The phase $\pf$ is a framing correction. If both
invariants are reduced to canonical framing, then
\qq
\pf=\frac{\pi}{12}\frac{K-c_V}{K}\dim G\,\left[\PU-\tspqn\right],
\label{1.11}
\qqq
here $\nu$ is a self-linking number of $\cK$ defined as a linking
number between $C_2$ and $\cK$.

For a more general case when a surgery is performed on a link $\cL$ in
$M$ Witten concluded that
\qq
\zmk{\prime}=e^{i\pf}\sum_{\ba_1,\ldots\ba_n\in\Delta_G}
Z_{\ba_1,\ldots,\ba_n}(M,\cL;k)\tilde{U}_{\ba_1\br}^{(p_1,q_1)}
\ldots\tilde{U}_{\ba_n\br}^{(p_n,q_n)}.
\label{1.12}
\qqq
Reshetikhin and Turaev showed in \cite{ReTu} that eq.~(\ref{1.12}) is
invariant under Kirby moves. Therefore they proved that $\zmk{}$ is a
topological invariant of the manifold without invoking the path
integral representation~(\ref{1.1}) which still lacks mathematical
rigor. They also established a general set of conditions on the
components of the r.h.s. of eq.~(\ref{1.12}) which guarantee its
topological invariance.

The disadvantage of eqs.~(\ref{1.10}) and (\ref{1.12}) is that they
do not make the relation between Witten's invariant and classical
topological invariants of 3d manifolds quite transparent
(Alexander polynomial was the
only quantum invariant which had a clear topological nature since it
was originally constructed from the fundamental group of the knot
complement).
A possible way to deal with this problem is to consider a
large $k$ asymptotics of the path integral~(\ref{1.1}) by applying
a stationary phase approximation. The stationary phase points
are flat connections. Therefore the invariant is presented as a sum
over connected pieces $\cM_c$ of the moduli space $\cM$ of flat
connections on $M$:
\begin{eqnarray}
Z_{\ba_1,\ldots,\ba_n}(M,\cL;k)&=&
\sum_{\cM_c}Z^{(\cM_c)}_{\ba_1,\ldots,\ba_n}(M,\cL;k),
\nonumber\\
Z^{(\cM_c)}_{\ba_1,\ldots,\ba_n}(M,\cL;k)&=&
%\sum_{\cM_c}
\exp\frac{i}{\hbar}\left(S_{CS}^{(c)}+\sum_{n=1}^{\infty}
S_n^{(c)}\hbar^n\right),
\label{1.13}
\end{eqnarray}
here $S_{CS}$ is a Chern-Simons action of flat connections of $\cM_c$
and $S_n^{(c)}$ are the quantum $n$-loop
corrections to the
contribution of $\cM_c$. The 1-loop correction is a determinant of the
quadratic form describing the small fluctuations of $S_{CS}(A_\mu)$
around a stationary phase point. Its major features were determined by
Witten \cite{Wi1}, Freed and Gompf \cite{FrGo}, and Jeffrey \cite{Je}
(some further details were added in \cite{Ro1}):
\begin{eqnarray}
\lefteqn{
e^{iS_1^{(c)}}=
\frac{
(2\pi\hbar)^{\frac{\dim H_c^0-\dim H^1_c}{2}}}
{\Vol(H_c)}
\exp\left(\frac{i}{2\pi}c_V S_{CS}-\frac{i\pi}{4}\Nph
\right)
}
\label{1.14}\\
\shift{\times
\int_{\cM_c}\left[
\sqrt{|\tau_R|}
\prod_{i=1}^{n}\thol{\ba_i}{L_i}\right],
}
\nonumber
%\label{1.14}
\end{eqnarray}
here $H_c$ is an isotropy group of $\cM_c$ (i.e. a subgroup of $G$
which commutes with the holonomies of connections $A_\mu^{(c)}$ of
$\cM_c$), $\Nph$ is expressed \cite{FrGo} as
\qq
\Nph=2I_c + \dim H_c^0 + \dim H_c^1 + (1+b_M^1)\dim G,
\label{1.014}
\qqq
$I_c$ is a spectral flow of the operator $L_-=\star D+D\star$ acting
on 1- and 3-forms, $D$ being a covariant derivative, $H^0_c$ and
$H^1_c$ are cohomologies of $D$, and $b^1_M$ is the first Betti
number of $M$. $\tau_R$ is a Reidemeister-Ray-Singer torsion. It was
observed in \cite{Je} that $\sqrt{\tau_R}$ defines a ratio of volume
forms on $\cM_c$ and $H_c$.

The higher loop corrections $S_n^{(c)}$ are calculated by Feynman
rules. They are expressed as multiple integrals of the products of
propagators taken over the manifold $M$ and the link $\cL$. Such
representation might make the nature of invariants $S_n^{(c)}$ more
transparent. Bar-Natan~\cite{BN} and Kontsevich~\cite{Ko1} studied the
Feynman diagrams related to the link. These diagrams produce Vassiliev
invariants. In particular, Bar-Natan observed that the 2-loop
correction to the $SU(2)$ invariant of the knot in $S^3$ is
proportional to the second derivative of its Alexander polynomial.

In their recent paper \cite{MeMo} Melvin and Morton
conjectured\footnote{
This conjecture was proven recently by D. Bar-Natan and S.
Garoufalidis~\cite{BNGa} at the level of weight systems.}
a rather
strict limitation
%
%\footnote{This conjecture
%was proven by Bar-Natan \cite{BNpc} at the level of Feynman
%diagrams.}
%
on the possible powers of $\a$ in the $K^{-1}$ expansion
of the $SU(2)$ Jones polynomial $\zas$
%
%.  They also suggested a
%
as well as a relation between the dominant part of this expansion and
the Alexander polynomial which generalizes the result of \cite{BN}.

The properties of Feynman diagrams related to the manifold were
studied in early papers \cite{ALR},\cite{GMM} and then by Axelrod and
Singer \cite{AxSi} and Kontsevich \cite{Ko2}. A convergence of those
diagrams was proven, however no multiloop diagrams were explicitly
calculated.  An ``experimental'' approach to their study was initiated
in \cite{FrGo} and \cite{Je}. Freed and Gompf checked the 1-loop
formula~(\ref{1.14}) by comparing it numerically to the surgery
formula~(\ref{1.12}) applied to some lens spaces and Seifert homology
spheres. L. Jeffrey transformed the surgery formula for lens spaces
and some mapping tori into the asymptotic form~(\ref{1.13}) thus
obtaining all the loop corrections for those manifolds. This program
was further extended to Seifert manifolds in \cite{Ro1}. It was
observed there among other things that the 2-loop correction to the
contribution of the trivial connection was proportional to
Casson-Walker invariant as calculated by C. Lescop \cite{Le1}.

In this paper we study the trivial connection contribution to Witten's
invariant of a knot, a link
and a manifold. In Section~\ref{*2} we prove the
relation between the Jones and Alexander polynomials of a knot
(Proposition~\ref{p2.1}) conjectured in \cite{MeMo} by relating the
former to the Reidemeister-Ray-Singer torsion of the knot complement.
We also generalize this result to the case of an arbitrary rational
homology sphere (\rhs).
In Section~\ref{*3} we derive a knot surgery
formula for the trivial
connection contribution to Witten's invariant
of a \rhs\  (Proposition~\ref{p3.1}). We show that at the 2-loop level
this formula coincides with Walker's formula \cite{Wa} for
Casson-Walker invariant.
This proves the relation between the 2-loop correction to the
contribution of the trivial connection and the Casson-Walker invariant
(Proposition~\ref{p3.2}) conjectured in \cite{Ro1}.
In Section~\ref{*4} we try to go
beyond \rhss\ by considering a Seifert manifold with nontrivial
rational homology. We derive a formula for the trivial connection
contribution to its Witten's invariant (Proposition~\ref{p4.3}) and
compare its properties to the partition function of a 2d gauge theory
studied by Witten \cite{Wi2}.
The results of Section~\ref{*2} are illustrated in Appendix,
where a large $k$
asymptotics of the Jones polynomial of a torus knot is calculated. The
contributions of reducible and irreducible connections in the knot complement
are identified. Similarly to the results of~\cite{Ro1}, the contribution
of the irreducible connections appears to be 2-loop exact.
%+++++++++++++++++++++++++++++++++++++++++++
\nsection{Jones Polynomial and Reidemeister-Ray-Singer
Torsion}
\label{*2}
%+++++++++++++++++++++++++++++++++++++++++++

We are going to study a Jones polynomial of a knot $\cK$ in a rational
homology sphere $M$ (i.e. $b_M^1=0$). We start with the case of
$M=S^3$. Then the $SU(2)$ Jones polynomial (in Witten's
normalization~(\ref{1.1})) can be expanded in $K^{-1}$:
\qq
\zas=\sum_{m,n\geq 0}C_{m,n}\a^m K^{-n}.
\label{2.1}
\qqq
Melvin and Morton \cite{MeMo} suggested\footnote{
I am thankful to D. Bar-Natan and S. Garoufalidis for drawing my
attention to the paper \cite{MeMo}.}
the following
\begin{proposition}
If the knot $\cK$ is canonically framed (i.e. the linking number $\nu$
between the cycle $C_2$ which determines
the framing and $\cK$ is zero),
then
\qq
C_{m,n}=0\;\;\;{\rm if}\;m>n.
\label{2.01}
\qqq
Moreover,
\qq
\sum_{n\geq 0}C_{n,n}a^n=\sqrt{\frac{2}{K}}\,
\frac{\sin(\pi a)}{\dask},\;\;0\leq a\leq 1,
\label{2.2}
\qqq
here $\dask$ is the Alexander polynomial of $\cK$ normalized in such a
way that $\Delta(S^3,{\rm unknot};\exp(2\pi i a))=1$, $\damk$ is real.
\label{p2.1}
\end{proposition}
It was established by Milnor \cite{Mi} and Turaev \cite{Tu} that in
this normalization $\Delta_A$ is related to the Reidemeister torsion
of the knot complement:
\qq
\damk=\frac{2\sin(\pi a)}{\trma}
\label{2.02}
\qqq
Some simple quantum field theory arguments were used in \cite{RoSa} to
show that the Alexander polynomial was related by eq.~(\ref{2.02}) to
the Ray-Singer torsion of the knot complement. Here we will apply the
same arguments to the Jones polynomial $\zas$.

Consider the values of $\a$ of order $K$. We introduce a new variable
\qq
a=\frac{\a}{K},\;\;0\leq a\leq 1.
\label{2.03}
\qqq
Let us split the path integral~(\ref{1.1}) for a knot $\cK$ into an
integral over the connection $A_\mu$ inside the tubular neighborhood
$\TK$ and inside its complement $S^3\setminus\TK$ with certain
boundary conditions on the boundary $T^2=\partial\TK$, as well as an
integral over these boundary conditions. According to \cite{EMSS}, one
possible set of boundary conditions requires that the gauge fields
$A_{1,2}$ on $T^2$ should belong to the Cartan subalgebra, the
curvature $F_{1,2}$ should be zero and the integral
$I_1=\oint_{C_1}A_\mu dx^\mu$ should be fixed. In fact, it was
established in \cite{EMSS} that in accordance with eq.~(\ref{1.5}),
the path integral over connections on $\TK$ is proportional to
$\delta(I_1-2\pi ia)$.
Therefore the Jones polynomial $\zas$ is equal to
the path integral over connections on $S^3\setminus\TK$
\qq
\zas=\int_{[S^3\setminus\TK]}[\cD A_\mu]
\exp\left(\frac{i}{\hbar}S_{CS}^\prime\right)
\label{2.4}
\qqq
taken with the boundary condition
\qq
\hol{C_1}=\exp(2\pi ia).
\label{2.5}
\qqq
The Chern-Simons action is modified \cite{EMSS} by the boundary term
\qq
S^\prime_{CS}=S_{CS}+\frac{1}{2}\Tr\int_{T^2}A_1A_2
\,d^2x,
\label{2.6}
\qqq
which is necessary for the choice~(\ref{2.5}) of boundary conditions.

Let us calculate the path
integral~(\ref{2.4}) by the stationary phase
approximation method~(\ref{1.13}). First of all, we look for
stationary phase
points, i.e. flat connections satisfying the boundary
condition~(\ref{2.5}). There is only one such connection for $a<a_0$
($a_0>0$ being a critical value depending on $\cK$). This connection
is reducible: all the holonomies belong to the maximal torus
$U(1)\subset SU(2)$. For this connection $S_{CS}=0$. Since the linking
number $\nu$ of $C_2$ and $\cK$ is zero, the homology class of $C_2$
in $S^3\setminus\TK$ is trivial. Therefore $A_2=0$ and the boundary
term in eq.~(\ref{2.6}) is also zero. Thus the whole classical
Chern-Simons action $S_{CS}^\prime$ is zero.

We will estimate the 1-loop correction~(\ref{1.14}) up to a phase
factor $\exp\left(-\frac{i\pi}{4}\Nph\right)$. The flat $U(1)$
connection on $S^3\setminus\TK$ satisfying eq.~(\ref{2.5}) has no
moduli, so $\dim H_c^1=0$. The isotropy group is $H_c=U(1)$, so $\Vol
H_c=2\sqrt{2}\pi$ (recall that the radius of $U(1)$ is $\sqrt{2}$),
while $\dim H_c^0=1$. The determinants in the $SU(2)$ Ray-Singer
torsion $\tau_R$ split into three factors for three Lie algebra
components of $A_\mu$ which have the definite $U(1)$ charge. The
chargeless Cartan subalgebra (i.e. diagonal) component of $A_\mu$
contributes $1$, while each of the two off-diagonal components
contribute the square root of the $U(1)$ torsion $\trsa$. As a result
of all this and eq.~(\ref{2.02}) we conclude that
%
%**********
\begin{proposition}
The loop expansion formula~(\ref{1.13}) for
the Jones polynomial of a knot
in $S^3$ can be presented in the form
\qq
\zas=\sqrt{\frac{2}{K}}\frac{\sin(\pi a)}{\dask}
\exp\left[i\sum_{n=1}^{\infty}
\left(\frac{2\pi}{K}\right)
S_{n+1}(a)\right],
\label{2.7}
\qqq
here $S_{n}(a)$ are the higher loop corrections for the path
integral~(\ref{2.4}), they depend on the boundary holonomy $\exp(2\pi
ia)$.
\label{p2.2}
\end{proposition}
%****************
%
We will show later that $e^{-i\Nph}=1$.

The substitution~(\ref{2.03}) turns
the r.h.s. of this equation into
the expansion~(\ref{2.1}) with limitations~(\ref{2.01}) and
property~(\ref{2.2}). We also learn that the sum of the terms
\qq
\sum_{m\geq 0}C_{m,m+n}\a^m K^{-m-n}
\label{2.8}
\qqq
comes from the $n$-loop Feynman diagrams (including disconnected ones)
in the knot complement $S^3\setminus\TK$.

Consider now a general \rhs\ $M$ with a knot $\cK$ inside it. This
time there may be many flat connections (both reducible and
irreducible) with a given holonomy~(\ref{2.5}) even if $a$ is very
small. Each of them will contribute to the stationary phase
approximation of the path integral~(\ref{2.4}) turning it into the
sum~(\ref{1.13}). We will concentrate on the reducible $U(1)$
connections because their 1-loop contributions can again be related to
the Alexander polynomial of $\cK$.

Some changes have to be made to eq.~(\ref{2.7}). Let $b$
define the holonomy along $C_2$ for a reducible flat connection on
$M\setminus\TK$:
\qq
\hol{C_2}=\exp(2\pi ib).
\label{2.9}
\qqq
The holonomies~(\ref{2.5}) and (\ref{2.9}) are related by the fact
that the homomorphism
\qq
H_1(\partial\TK,\ZZ)\rightarrow H_1(M\setminus\TK,\ZZ)
\label{2.10}
\qqq
has a kernel. Let the cycle
\qq
C_0=d\,(m_1 C_1 +
m_2 C_2),\;\;\;\;d,m_1,m_2\in\ZZ,\;\;m_1,m_2\;-\;{\rm
coprime}
\label{2.11}
\qqq
be its generator. Then
\qq
\hol{C_0}\equiv\exp[2\pi id(m_1 a + m_2 b)]=1,
\label{2.011}
\qqq
so that
\qq
b=-\frac{1}{m_2}\left(m_1 a + \frac{n}{d}\right),\;\;\;n\in\ZZ,\;\;
0\leq n<d.
\label{2.12}
\qqq
If we smoothly reduce $a$ to zero, then the flat connection on
$M\setminus\TK$ will transform into a flat connection on $M$. Let
$S_{CS,0}$ be its Chern-Simons invariant. Then according to
\cite{KiKl} and eq.~(\ref{2.6}), the Chern-Simons action of the
original connection is
\qq
S_{CS}^\prime=-\pi^2
\left(\frac{m_1}{m_2}a^2 + 2\frac{na}{m_2 b}\right) +
S_{CS,0}.
\label{2.13}
\qqq
In particular, if the flat connection on $M$ at $a=0$ is trivial, then
$S_{CS}=0$ and $n=0$, so that
\qq
S_{CS}^\prime= -\pi^2 \frac{m_1}{m_2}a^2.
\label{2.1013}
\qqq
The Reidemeister-Ray-Singer torsion for the Cartan subalgebra part of
$A_\mu$ is known to be equal to $\ohm$. As for the off-diagonal
Lie algebra components of $A_\mu$, we can use again eq.~(\ref{2.02}).
However the argument of the torsion is related to the holonomy along
the generator of the $\ZZ$ part of $H_1(M\setminus\TK,\ZZ)$. The
holonomy along the cycle $C_0^\prime$ which has the unit intersection
number with $C_0$, is
\qq
\hol{C^\prime_0}=\exp\left(2\pi i\frac{a}{m_2}\right).
\label{2.013}
\qqq
This cycle generates the $d\ZZ$ subgroup of $\ZZ$, so the holonomy
along the generator of $\ZZ$ is $\eamd$. Combining all pieces together
we get the following
%
%********************
\begin{proposition}
If $M$ is a \rhs\ and $\cK$ is a knot inside it, then the contribution
of the trivial connection to the Jones polynomial of $\cK$ is
given by the formula
\begin{eqnarray}
\ztramk &=&\sqrt{\frac{2}{K}}\,[\ohm]^{-\frac{1}{2}}
\frac{\sin\left(\frac{\pi}{K}\frac{\a}{m_2 d}\right)}
{\Delta_A\left(M,\cK,\exp\left(\frac{2\pi i}{K}\frac{\a}{m_2 d}
\right)\right)}
\exp\left(-\frac{i\pi}{2K}\frac{m_1}{m_2}\a^2\right)
\nonumber\\
&&\times\exp\left[
\sum_{n=1}^{\infty}\left(\frac{2\pi}{K}\right)^n
S_{n+1}\left(\frac{\a}{K}\right)\right].
\label{2.14}
\end{eqnarray}
\end{proposition}
We dropped the factor $e^{\frac{i\pi}{4}\Nph}$, we will show later
that it is equal to 1.

Assuming that $\a\ll K$ (that is, $a\ll 1$), we can present $\ztramk$
in a slightly different form by applying the stationary phase
approximation directly to the path integral~(\ref{1.1}) taken over
connections on the whole manifold $M$:
\qq
\ztramk =\ztrmk\,\exp\left[\frac{i\pi}{2K}\nu(\a^2-1)\right]
\,\a J(\a,K).
\label{2.15}
\qqq
In this formula $\ztrmk$ is a contribution of the trivial connection
to Witten's invariant of $M$ itself, it contains Feynman diagrams
which are not connected to the knot $\cK$.
The function $J(\a,K)$ is a contribution of Feynman diagrams attached
to the knot, except for two factors that we separated out explicitly:
the framing factor $\exp\left[\frac{i\pi}{2K}\nu(\a^2-1)\right]$
and the dimension of the representation $\dim V_\a=\a$, which appears
when the trace of the holonomy is taken in eq.~(\ref{1.1}).
%
%The remaining factors come
%from Feynman diagrams attached to the knot. We separated explicitly
%the framing factor $\exp\left[\frac{i\pi}{2K}\nu(\a^2-1)\right]$ and
%the dimension of the representation $\dim V_\a=\a$, which appears
%when the trace of the holonomy is taken in eq.~(\ref{1.1}).
%
Note the
relation between the self-linking number of the knot $\nu$ and the
numbers $m_1,m_2$:
\qq
\nu=-\frac{m_1}{m_2}.
\label{2.016}
\qqq

The function $J(\a,K)$ can be expanded in $K^{-1}$
\qq
J(\a,K)=\sum_{m,n\geq 0}D_{m,n}\a^m K^{-n}.
\label{2.16}
\qqq
The numbers $D_{m,n}$ are type $n$ Vassiliev invariants.
By comparing eqs.~(\ref{2.14}) and (\ref{2.15}) we conclude that
\qq
D_{m,n}=0\;\;\;{\rm for}\;\;m>n.
\label{2.17}
\qqq
Moreover, according to \cite{BN} $D_{0,0}=1$,
$D_{0,1}=D_{1,1}=D_{1,2}=0$, $D_{2,2}=-D_{2,0}$, so that
\qq
J(\a,K)=1+D_{2,2}\frac{\a^2 -1}{K^2}+\cO(K^3).
\label{2.18}
\qqq
The value of $D_{2,2}$ can be deduced by comparing eqs.~(\ref{2.14})
and (\ref{2.15}):
\qq
D_{2,2}=\frac{2\pi^2}{m_2^2 d^2}\left(\Delta_A^{\prime\prime}
-\frac{1}{12}\right)
\label{2.018}
\qqq
This relation was first obtained by Bar-Natan in \cite{BN}.

The trivial connection contribution $\ztrmk$ can be expanded in
asymptotic series in $K^{-1}$. The leading 1-loop term is given by
eq.~(\ref{1.14}):
\qq
Z_{\rm 1-loop}^{({\rm tr})}(M;k)=\frac{1}{\Vol SU(2)}
\left(\frac{2\pi\hbar}{\ohm}\right)^{\frac{3}{2}}\equiv
\sqrt{2}\pi(K\,\ohm)^{-\frac{3}{2}}.
\label{2.1018}
\qqq
Comparing eqs.~(\ref{2.1018}), (\ref{2.15}) and (\ref{2.14}) we see
that $\Nph=0$ and the term $e^{-\frac{i\pi}{4}\Nph}$ can indeed be
dropped from eqs.~(\ref{2.14}) and (\ref{2.7}).

All the formulas of this section can be easily generalized to the case
of a general simple Lie group $G$. Equation~(\ref{2.14}) transforms
into
\begin{eqnarray}
\ztrbamk&=&[2K\,\ohm]^{-\frac{\rank G}{2}}
\exp\left(-\frac{i\pi}{K}\frac{m_1}{m_2}\a^2\right)
\label{2.19}\\
&&\times
\left[\prod_{\bl_i\in\Delta_+}\frac
{2\sin\left(\frac{\pi}{K}\frac{\ba\cdot\bl_i}{m_2 d}\right)}
{\Delta_A\left(M,\cK;\exp\left(\frac{2\pi i}{K}\frac
{\ba\cdot\bl_i}{m_2 d}\right)\right)}\right]
\exp\left[i\sum_{n=1}^{\infty}\left(\frac{2\pi}{K}\right)
^{-n}S_{n+1}
\left(\frac{\ba}{K}\right)\right].
\nonumber
%\label{2.19}
\end{eqnarray}
The generalization of eqs.~(\ref{2.15}) and (\ref{2.18}) is
\begin{eqnarray}
\ztrbamk&=&\ztrmk\exp\left(\frac{i\pi}{K}\nu(\ba^2 -\br^2)\right)
\left[\prod_{\bl_i\in\Delta_+}
\frac{(\ba\cdot\bl_i)}
{(\br\cdot\bl_i)}\right]\,J(\ba,K),
\label{2.20}\\
J(\ba,K)&=&1+D_{2,2}\frac{12\br^2}{\dim G}
\frac{\ba^2-\br^2}{K^2} +\cO(K^{-3}).
\label{2.21}
\end{eqnarray}
%

%*********************************
\nsection{A Trivial Connection Contribution to Witten's Invariant}
\label{*3}
%*********************************

Suppose that a manifold $M^\prime$ is constructed by a $\Upq$ surgery
on a knot $\cK$ in a manifold $M$. Then Witten's invariant of
$M^\prime$ can be calculated by the surgery formula~(\ref{1.10}). The
large $k$ limit of the r.h.s. of this formula contains implicitly the
contributions of all flat connections on $M^\prime$. We will try to
separate the contribution of the trivial connection in the case when
both $M$ and $M^\prime$ are rational homology spheres.

We start with the case of $G=SU(2)$. Our main tool is the finite
version of the Poisson resummation formula. The Poisson resummation
formula states that for any function $f(\a)$
\qq
\sum_{\a\in\ZZ}f(\a)
=\sum_{l\in\ZZ}\int d\a \exp(2\pi i\a l)\,f(\a).
\label{3.1}
\qqq
Therefore we would like to extend the sum in eq.~(\ref{1.10}) from
$1\leq\a\leq K-1$ to $\ZZ$. First of all, since $\tu$ is equal to zero
at $\a=0,K$ we can add these points to the range of summation. We can
also double this range:
\qq
\sum_{\a=0}^{K}\longrightarrow \frac{1}{2}\sum_{\a=-K+1}^{K}
\label{3.01}
\qqq
because the summand is even. Finally, we use a ``regularization''
formula
\qq
\sum_{\a=-K+1}^{K}f(\a)=2K\lim_{\epsilon\rightarrow 0}
\epsilon^{\frac{1}{2}}\sum_{\a\in\ZZ}\ep f(\a),\;\;\;{\rm if}\;\;
f(\a+2K)=f(\a).
\label{3.2}
\qqq
Thus we obtain a sum $\sum_{\a\in\ZZ}$ to which we apply
eq.~(\ref{3.1}):
\qq
\zmk{\prime}=e^{i\pf} K\lime\epsilon^{\frac{1}{2}}
\sum_{l\in\ZZ}\int_{-\infty}^{+\infty}d\a\,\ep\exp(2\pi il\a)\,
\zamk\,\tu.
\label{3.3}
\qqq

At this point we make the following assumption: the large $k$ limit of
the integral over $\a$ in eq.~(\ref{3.3}) is equal to the sum of
the contributions of the special points $\a_*$ of the integrand, e.g.
stationary phase points, breaks, poles, \etc. Their contribution
depends on the local properties of the integrand. Apart from the
regularization factor $\ep$, the set of critical points for all
$l\in\ZZ$ and their respective contributions exhibit the same
symmetries as the original summand of eq.~(\ref{1.10}), i.e. they are
even and have a period of $2K$. The role of the factor $\ep$ to the
leading power in $\epsilon$ is to suppress the contribution of each
critical point $\a_*$ by a factor $e^{-\e\pi\a_*^2}$. Therefore we can
play eq.~(\ref{3.2}) backwards: we drop $K\lime \e^{\frac{1}{2}}$ and
$\ep$ while limiting ourselves to the contribution of only those
critical points which belong to the fundamental domain
\qq
0\leq\a_*\leq K.
\label{3.4}
\qqq
In other words,
\qq
\zmk{\prime}=e^{i\pf}\sum_{l\in\ZZ}
\int^{+\infty}_{\stackrel{-\infty}{[0\leq\a_*\leq K]}}
d\a\,e^{2\pi il\a}\zamk\,\tu,
\label{3.03}
\qqq
here the symbol
$\int^{+\infty}_{\stackrel{-\infty}{[0\leq\a_*\leq K]}}$
means that we take only the contributions of the special
points~(\ref{3.4}). If $\a_*=0,K$, then its contribution to the
integral of eq.~(\ref{3.03}) carries an extra boundary
factor~$\frac{1}{2}$.

We assume that each of the special points in the domain~(\ref{3.4})
corresponds to one or several connected pieces
$\cM_c^\prime$ of the moduli
space
$\cM^\prime$ of flat connections on
$M^\prime$. Consider a cycle in $M^\prime$
which corresponds to the cycle $C_2$ on the boundary of a tubular
neighborhood of the knot $\cK$ in $M$. We will also call it
$C_2$.
According to eq.~(\ref{1.5}),
the holonomy of a flat connection related to a special point
$\a_*$ along $C_2$ is equal to $\east$, so
the contribution of the trivial connection on $M^\prime$ should come
from the point $\a_*=0$.

We now concentrate on the point $\a_*=0$, so we are interested only in
the values $\a\ll K$. Therefore we can use  a 1-loop approximation
given by  the formulas~(\ref{1.13}) and (\ref{1.14}) for the partition
function $\zamk$. Some of the terms $Z^{(\cM_c)}_\a(M,\cK;k)$ may have
a special point $\a_*=0$. We have to determine which of them do
contribute to $\ztrmpk$.

We are going to present some arguments why the contribution to
$\ztrmpk$ may actually come only from $\ztramk$. Suppose that a
contribution of the special point $\a_*=0$ of a particular term
$Z^{(\cM_c)}_\a(M,\cK;k)$ to the r.h.s. of eq.~(\ref{3.03})
corresponds to the contribution $\ztrmpk$ of the trivial
connection to
the r.h.s. of that equation. Let us multiply the integrand of the path
integral for $\zmk{\prime}$ by an ``observable'' factor
\qq
\cO(C,\b)=\thol{\b}{C}
\label{3.0*1}
\qqq
for $\b\ll K$, thus turning it into
$Z_\b(M^\prime,C;k)$.
According to eq.~(\ref{1.14}) in the 1-loop approximation
\qq
Z^{({\rm tr})}_\b(M^\prime,C;k)=\b\ztrmpk.
\label{3.0*2}
\qqq
A surgery formula~(\ref{3.03}) can also work for $Z_\b(M^\prime,C;k)$
if we add the factor~(\ref{3.0*1}) to $\zamk$ transforming it into
$Z_{\a\b}(M,(\cK,C);k)$. According to eq.~(\ref{1.14}), the effect of
the factor $\cO(C,\b)$ on
$Z_\a^{(\cM_c)}(M,\cK;k)$
is (if we forget for a moment about an integral over $\cM_c$) to
multiply it by a factor $\Tr_\b\Pexp\left(\oint_C
A_\mu^{(c)}dx^\mu\right)$.  This factor turns into $\b$ if $\cM_c$ is
a trivial connection.  This is not only a sufficient but also a
necessary condition if $\cM_c$ is a point.

If $\cM_c$ has a nonzero dimension and there is  a
nontrivial integral in eq.~(\ref{1.14}) then we may use the following
reasoning. Characters form a basis in the space of functions on the
maximal torus of the Lie group (in our case it is $U(1)\subset
SU(2)$). Therefore we can take a linear combination of observables
with different values of $\b$ so that they form a smooth slowly
varying function on the space of conjugation classes of holonomies
$\hol{C}$, which is equal to 1 at identity. This new observable is
equal to 1 on $\ztrmpk$ (that is, if we multiply the integrand of
the path integral for $\ztrmpk$ by that observable, then the value of
the path integral does not change at 1-loop). However different
choices of the smooth function will affect the value of the integral
over $\cM_c$ in eq.~(\ref{1.14}).
Therefore we conclude that the contribution of the trivial connection
on $M^\prime$ to $\zmk{\prime}$ comes only from the contribution
$\ztramk$ of the trivial connection on $M$ to $\zamk$.

According to eq.~(\ref{2.15}), if $M$ is a \rhs\, then $\a=0$ is not
a singular point of $\ztramk$. Therefore its only chance to contribute
to the integral~(\ref{3.03}) is to be a stationary phase point.
According to eqs.~(\ref{1.9}), (\ref{2.14}) and
(\ref{2.15}), the relevant part of
the phase is
\qq
\frac{i\pi}{2K}\pqn\a^2+2\pi i\left(l+\frac{n}{q}\right)\a.
\label{3.*04}
\qqq
We see that $\a=0$ can indeed be a stationary phase point if we put
$n=0$ in eq.~(\ref{1.9}) and $l=0$ in eq.~(\ref{3.03}). Now it remains
to substitute eqs.~(\ref{1.9}) and (\ref{2.15}) into the r.h.s. of
eq.~(\ref{3.03}) and add an extra boundary factor $\frac{1}{2}$. Then
we come to the following
%
%****************
\begin{proposition}
If $M$ and $M^\prime$ are rational homology spheres and $M^\prime$ is
constructed by a rational surgery $\Upq$ on a knot $\cK$ in $M$, which
has a self-linking number $\nu$, then the trivial connection
contribution to Witten's invariants of $M$ and $M^\prime$ are related
by the formula
\begin{eqnarray}
\lefteqn{\ztrmpk=}
\label{3.*4}\\
&&=\ztrmk
%\ztrmpk&=&\ztrmk\,
%\frac{1}{2}
\frac{\sign{q}}{\sqrt{2K|q|}}
e^{-i\frac{3}{4}\pi\sign{\frac{p}{q}+\nu}}
\exp\left[\frac{i\pi}{2K}\left(12s(p,q)-\pqn+\tspqn\right)\right]
\nonumber\\
\shift{
\times\int^{+\infty}_{\stackrel{-\infty}{[\a_*=0]}}
d\a\,\sin\left(\frac{\pi\a}{Kq}\right)\,\a\,J(\a,K)\,
\exp\left[\frac{i\pi}{2K}\pqn\a^2\right],
}
\nonumber
%\label{3.*4}
\end{eqnarray}
here the function $J(\a,K)$ comes from eq.~(\ref{2.15}), it is a
Feynman diagram contribution of the trivial connection to the Jones
polynomial of $\cK$ and satisfies the
properties~(\ref{2.16})--(\ref{2.018}).

The integral
$\int^{+\infty}_{\stackrel{-\infty}{[\a_*=0]}}$ in eq.~(\ref{3.4})
should be calculated in the following way: the preexponential factor
$\sin\left(\frac{\pi\a}{Kq}\right)\,\a\,J(\a,K)$ should be expanded in
$K^{-1}$ series with the help of eq.~(\ref{2.16}), then each term
should be integrated separately with the gaussian factor
$\exp\left[\frac{i\pi}{2K}\pqn\right]$.
\label{p3.1}
\end{proposition}
%*********************
%

According to this prescription a term $D_{m,n}\a^M K^{-n}$ in the
expansion~(\ref{2.16}) contributes up to the
$\left(n-\frac{m}{2}\right){\rm th}$ order in the loop expansion of
$\ztrmpk$. Therefore the limitation~(\ref{2.17}) leads to the
following
%
%***************
\begin{corollary}
Only a finite number of Vassiliev's invariants participate in a
surgery formula for $\ztrmpk$ at a given loop order.
\label{c3.1}
\end{corollary}
%*****************
%
In particular, we present explicit surgery formulas for the first two
loop corrections. In the notations of eq.~(\ref{1.13})
\begin{eqnarray}
\lefteqn{
e^{iS_1^{({\rm tr})}(M^\prime)}=
|p+\nu q|^{-\frac{3}{2}}\,e^{iS_1^{({\rm tr})}(M)}
}
\label{3.5}\\
\lefteqn{
S_2^{({\rm tr})}(M^\prime)=
S_2^{({\rm tr})}(M)+3\left[s(p,q)-\frac{1}{12}\pqn+
\frac{1}{4}\spqn
\right.
}
\label{3.6}
\\
\shift{
\left.
+\frac{1}{\pqn}\left(\frac{D_{2,2}}{2\pi^2}-
\frac{1}{12}\frac{1}{q^2}\right)\right].
}
\nonumber
\end{eqnarray}
The first formula is consistent with eq.~(\ref{1.14}) which predicts
that for a \rhs\
\qq
e^{iS_1^{({\rm tr})}(M)}=\sqrt{2}\pi [K\ohm]^{-\frac{3}{2}}.
\label{3.7}
\qqq
As for eq.~(\ref{3.6}), it transforms into Walker's surgery formula
\cite{Wa} for Casson's invariant of a \rhs\footnote{
I am thankful to K. Walker for checking this.}\
if we substitute
\qq
S_2^{({\rm tr})}(M)=3\l_{CW}(M)
\label{3.8}
\qqq
and recall the relation~(\ref{2.018}) between $D_{2,2}$ and the second
derivative of the Alexander polynomial. Thus we conclude that the
conjecture of \cite{Ro1} can indeed be extended to all \rhss:
%
%************************
\begin{proposition}
If $M$ is a rational homology sphere, then the 2-loop correction to
the contribution of the trivial connection to its Witten's invariant
(defined by eq.~(\ref{1.13})) is related to the Casson-Walker
invariant according to eq.~(\ref{3.8}).
\label{p3.2}
\end{proposition}
%*************************
%
In the case of a general simply laced Lie group $G$ the surgery
formula~(\ref{3.*4}) takes the form
\begin{eqnarray}
\ztrmpk&=&\ztrmk\frac{[2\sign{q}]^{|\Delta_+|}}
{(K|q|)^{\frac{\rank G}{2}}}\frac{1}{|W|}
e^{-\frac{i\pi}{4}\dim G\,\spqn}
\label{3.9}
\\
&&\times
\exp\left[\frac{i\pi}{K}\br^2\left(12s(p,q)-\pqn+\tspqn\right)\right]
\nonumber\\
&&\times\int_{[\ba_*=0]}d\ba
\left[\prod_{\bl_i\in\Delta_+}
\frac{(\ba\cdot\bl_i)}{(\br\cdot\bl_i)}
\sin\left(\frac{\pi}{Kq}(\ba\cdot\bl_i)\right)\right]
\nonumber\\
&&\hspace*{2in}\times
J(\ba,K)\exp\left[\frac{i\pi}{K}\pqn\ba^2\right],
\nonumber
%\label{3.9}
\end{eqnarray}
here $|W|$ is the number of elements in the Weyl group and the
integral goes over the Cartan subalgebra. The first two loop
corrections are
\begin{eqnarray}
e^{iS_1^{({\rm tr})}(M)}&=&
\frac{1}{\Vol G}\left[\frac{2\pi}{K}\frac{1}{\ohm}\right]^{
\frac{\dim G}{2}},
\label{3.10}\\
S_2^{({\rm tr})}(M)&=&6\br^2\l_{CW}(M).
\label{3.11}
\end{eqnarray}
A simple formula
\begin{eqnarray}
\lefteqn{
\int d\a\exp\left[\frac{i\pi}{K}\pqn\ba^2\right]
\prod_{\bl_i\in\Delta_+}
\frac{(\ba\cdot\bl_i)}{(\br\cdot\bl_i)}
\sin\left[\frac{\pi}{Kq}(\ba\cdot\bl_i)\right]
}
\label{3.12}
\\
\shift{
=e^{\frac{i\pi}{4}\dim G\,\spqn}|W|
\left[\frac{\sign{q}}{2|q|}\right]^{|\Delta_+|}
K^{\frac{\rank G}{2}}\left|\frac{p}{q}+\nu\right|^{-\frac{\dim G}{2}}
\exp\left(-\frac{i\pi}{K}\frac{\br^2}{q(pq+\nu)}\right)
}
\nonumber
\end{eqnarray}
allows an easy check of an obvious generalization of
relations~(\ref{3.5}) and (\ref{3.6}).

%*********************************
\nsection{Beyond the Rational Homology Spheres}
\label{*4}
%*********************************
If a manifold $M$ ($M^\prime$) is not a \rhs\ then the trivial
connection is a point on a connected piece $\cM_0$ ($\cM_0^\prime$) of
the moduli space of flat connections. Equations~(\ref{2.15}),
(\ref{2.16}) and (\ref{2.18}) are no longer valid, since the 1-loop
contribution of $\cM_0$ to the partition function $\zamk$ includes an
integral
\qq
\int_{\cM_0}\sqrt{|\tau_R|}\thol{\a}{\cK}
\label{4.1}
\qqq
which may have singularities (e.g. poles or breaks) at $\a=0$.

We will determine the contribution of $\cM_0^\prime$ to Witten's
invariant $\zmk{\prime}$ when $M^\prime$ is a Seifert manifold
$\xgf$. $X_g$ can be produced by $n$ surgeries $U^{(p_i,q_i)}$ on
fibers of $S^1\times\Sigma_g$, $\Sigma_g$ is a $g$-handle Riemann
surface. We will sketch the calculation leaving the details for
\cite{Ro2}.

The Seifert manifold $\xgf$ can be constructed by an $S$
surgery\footnote{$S=\left(\begin{array}{cc}
0&1\\-1&0\end{array}\right)\in
SL(2,\ZZ)$.} on a special knot $\cK$ belonging to the manifold $M$
which is a connected sum of $n$ lens spaces $L_{-p_i,q_i}$ and $2g$
manifolds $S^1\times T^2$. The Jones polynomial of the knot $\cK$ (in
Witten's normalization) is
\qq
\zamk=\prod_{i=1}^{n}\tilde{U}_{\a 1}^{(-q_i,p_i)}
\tilde{S}_{\a 1}^{1-2g-n}.
\label{4.2}
\qqq
We put $n=0$ in eq.~(\ref{1.9}) in order to extract the contribution
of $\cM_0$ to $\zamk$:
\begin{eqnarray}
\lefteqn{
Z^{(\cM_0)}_\a(M,\cK;k)=
}
\label{4.3}\\
&&=
\left(\frac{2}{K}\right)^{\frac{1}{2}-g}
\exp\left(-\frac{i\pi}{2K}\frac{H}{P}\a^2\right)
\left[\prod_{j=1}^{n}\frac{\sign{p_j}}{\sqrt{|p_j|}}
e^{-\frac{i\pi}{4}\PUm}\right]
\frac{\prod_{j=1}^{n}\sin\left(\frac{\pi}{K}\frac{\a}{p_j}\right)}
{\sin^{n+2g-1}\left(\frac{\pi}{K}\a\right)},
\nonumber
\end{eqnarray}
here
\qq
P=\prod_{j=1}^{n}p_j,\;\;\frac{H}{P}=\sum_{j=1}^{n}\frac{q_j}{p_j},\;\;
H=\ord H_1(X_g,\ZZ).
\label{4.4}
\qqq
The expression in the r.h.s. of eq.~(\ref{4.3}) has a pole at $\a=0$
of order $2g-1$.

After substituting eq.~(\ref{4.3}) into the surgery
formula~(\ref{1.10}) and taking into account that $\Upq=S$ and
\qq
\pf=\frac{\pi}{4}\frac{K-2}{K}\left[\sum_{j=1}^{n}\PUm+
3\sign{\frac{H}{P}}\right]
\label{4.5}
\qqq
we get the following expression
\qq
\left(\frac{2}{K}\right)^{1-g}\sum_{\a=1}^{K-1}
\exp\left(-\frac{i\pi}{2K}\frac{H}{P}\a^2\right)
\frac{\sign{P}}{\sqrt{|P|}}
\frac{e^{i\pi\frac{3}{4}\frac{K-2}{K}\sign{\frac{H}{P}}}}
{\sin^{n+2g-2}\left(\frac{\pi}{K}\a\right)}
\prod_{j=1}^{n}e^{-\frac{i\pi}{2K}\PUm}\sin\left(
\frac{\pi}{K}\frac{\a}{p_j}\right).
\label{4.6}
\qqq
We should be careful in converting it to the integral~(\ref{3.03}),
because the points $\a=0,K$ are singular. We shift the argument of the
singular factor:
\qq
\sin^{-n-2g+2}\left(\frac{\pi}{K}\a\right)
\longrightarrow \lime\sin^{-n-2g+2}\left[
\frac{\pi}{K}(\a-i\e)\right].
\label{4.7}
\qqq
Now we can add the points $\a=0,K$ to the sum~(\ref{4.6}). The factor
$\prod_{j=1}^{n}\sin\left(\frac{\pi}{K}\frac{\a}{p_j}\right)$ makes the
contribution of $\a=0$ equal to zero, while the contribution of $\a=K$
does not affect the local behavior at $\a=0$. Thus the contribution of
the piece $\cM_0^\prime$ containing the trivial connection to Witten's
invariant $Z(X_g;k)$ is equal to the contribution of the special
point $\a_*=0$ to the integral in the following expression
\begin{eqnarray}
Z^{(\cM_0^\prime)}(X_g;k)&=&
\frac{1}{2}\left(\frac{2}{K}\right)^{1-g}\frac{\sign{P}}{\sqrt{|P|}}
e^{i\pi\frac{3}{4}\sign{\frac{H}{P}}}
\exp\frac{i\pi}{2K}\left(\frac{H}{P}-3\sign{\frac{H}{P}}
-12\sum_{j=1}^{n}s(q_j,p_j)\right)
\nonumber\\
&&\times\lime
\sum_{l\in\ZZ}\int^{+\infty}_{\stackrel{-\infty}{\a_*=0}}
d\a\,\frac{\prod_{j=1}^{n}\sin\left(\frac{\pi}{K}\frac{\a}{p_j}\right)}
{\sin^{n+2g-2}\left(\frac{\pi}{K}(\a-i\e)\right)}
e^{2\pi i\a l}\exp\left(-\frac{i\pi}{2K}\frac{H}{P}\a^{2}\right).
\label{4.8}
\end{eqnarray}

For $l=0$, $\a=0$ is a stationary phase point. Similar to the previous
section we conclude that
%
%*****************
\begin{proposition}
The contribution of the $l=0$ term to the expression~(\ref{4.8}) for
$Z^{(\cM_0^\prime)}(X_g;K)$ is equal to
\begin{eqnarray}
Z^{(\cM_0^\prime)}_{l=0}&=&
\frac{1}{2}\left(\frac{2}{K}\right)^{1-g}\frac{\sign{P}}{\sqrt{|P|}}
e^{i\pi\frac{3}{4}\sign{\frac{H}{P}}}
\exp\frac{i\pi}{2K}\left(\frac{H}{P}-3\sign{\frac{H}{P}}
-12\sum_{j=1}^{n}s(q_j,p_j)\right)
\nonumber\\
&&\times
\int^{+\infty}_{\stackrel{-\infty}{\a_*=0}}
d\a\,\sin^{-n-2g+2}\left(\frac{\pi}{K}\a\right)\,
\left[\prod_{j=1}^{n}\sin\left(\frac{\pi}{K}\frac{\a}{p_j}\right)
\right]\,\exp\left(-\frac{i\pi}{2K}\frac{H}{P}\a^2\right).
\label{4.08}
\end{eqnarray}
To calculate the integral we have to expand the preexponential factor
in Laurent series in $K^{-1}$:
\qq
\frac{\prod_{j=1}^{n}\sin\left(\frac{\pi}{K}\frac{\a}{p_j}\right)}
{\sin^{n+2g-2}\left(\frac{\pi}{K}(\a-i\e)\right)}
=\frac{1}{P}\left(\frac{\pi}{K}\a\right)^{2g-2}-
\frac{1}{6P}\left(\frac{\pi}{K}\a\right)^{4-2g}
\left(2-2g-n+\sum_{j=1}^{n}\frac{1}{p_j^2}\right)
+\ldots
\label{4.9}
\qqq
and then integrate each term separately with the help of the formula
\qq
\int_{-\infty}^{+\infty}d\a\,\a^{2m}
\exp\left(-\frac{i\pi}{2K}\frac{H}{P}\a^2\right)=
\left(\frac{2K}{\pi}\left|\frac{P}{H}\right|\right)^{m+\frac{1}{2}}
e^{-\frac{i\pi}{2}\left(m+\frac{1}{2}\right)
\sign{\frac{H}{P}}}\,\Gamma\left(m+\frac{1}{2}\right),
\label{4.10}
\qqq
which works for both positive and negative $m$.
\label{p4.1}
\end{proposition}
%********************
%
Thus the contribution of $l=0$ is expressed as an asymptotic series in
$K^{-1}$:
\begin{eqnarray}
Z_{l=0}^{(\cM_0^\prime)}&=&
2^{\frac{3}{2}-2g}\pi^{\frac{1}{2}-g}K^{2g-\frac{3}{2}}
|H|^{-\frac{3}{2}}\left(\frac{H}{P}\right)^{g}
\Gamma\left(\frac{3}{2}-g\right)
\left\{1+\right.
\nonumber\\
&&
+\frac{i\pi}{2K}\left[\frac{P}{H}
\left(1-\frac{2}{3}g\right)\left(2-2g-n+
\sum_{j=1}^{n}\frac{1}{p_j^2}\right)
-3\sign{\frac{H}{P}}+\frac{H}{P}
-12\sum_{j=1}^{n}s(q_j,p_j)\right]
\nonumber\\
&&\left.
+\cO(K^{-2})\right\}.
\label{4.09}
\end{eqnarray}

The stationary phase points of the integral in eq.~(\ref{4.8}) for
$l\neq 0$ are
\qq
\a^{({\rm st})}_l=2K\frac{H}{P}l.
\label{4.010}
\qqq
The integration contour of steepest descent for these points is
\qq
{\rm Im}\,\a=\sign{\frac{H}{P}}(\a^{({\rm st})}_l
-{\rm Re}\,\a).
\label{4.11}
\qqq
The original integration contour ${\rm Im}\,\a=0$ should be rotated
into the contour~(\ref{4.11}).
During this rotation it crosses the pole
$\a=i\e$ of the integrand of (\ref{4.8}) if
$\sign{\frac{H}{P}}\a^{({\rm st})}_l>0$. Therefore the pole
contributes its residue to the integral~(\ref{4.8}) if
\qq
l\geq 1.
\label{4.12}
\qqq
The total contribution of $l\neq 0$ to $Z^{(\cM_0^\prime)}(X_g;k)$ is
equal to
\begin{eqnarray}
Z^{(\cM_0^\prime)}_{l\neq 0}(X_g;k)&=&
\frac{1}{2}\left(\frac{2}{K}\right)^{1-g}\frac{\sign{P}}{\sqrt{|P|}}
e^{i\pi\frac{3}{4}\sign{\frac{H}{P}}}
\exp\frac{i\pi}{2K}\left(\frac{H}{P}-3\sign{\frac{H}{P}}
-12\sum_{j=1}^{n}s(q_j,p_j)\right)
\nonumber\\
&&\times 2\pi i\lime\Res_{\a=i\e}\left[
\frac{\exp\left(-\frac{i\pi}{2K}\frac{H}{P}\a^2\right)}
{e^{-i\pi \a}-1}
\frac{\prod_{j=1}^{n}\sin\left(\frac{\pi}{K}\frac{\a}{p_j}\right)}
{\sin^{n+2g-2}\left(\frac{\pi}{K}(\a-i\e)\right)}\right]
\label{4.009}
\end{eqnarray}
The $\e\rightarrow 0$ limit in this expression is nonsingular, so
after changing the variable to $x=\frac{\pi}{K}\a$, we conclude that
%
% *************************
\begin{proposition}
The contribution of $l\neq 0$ terms to the expression~(\ref{4.8}) for
$Z^{(\cM_0^\prime)}(X_g;K)$ is equal to
\begin{eqnarray}
Z^{(\cM_0^\prime)}_{l\neq 0}(X_g;k)&=&
-i\pi\left(\frac{2}{K}\right)^{1-g}\frac{\sign{P}}{\sqrt{|P|}}
e^{i\pi\frac{3}{4}\sign{\frac{H}{P}}}
\label{4.*09}
\\
&&\times
\exp\frac{i\pi}{2K}\left(\frac{H}{P}-3\sign{\frac{H}{P}}
-12\sum_{j=1}^{n}s(q_j,p_j)\right)
\nonumber\\
&&\times\Res_{x=0}\frac{K}{\pi}
\frac{\exp\left(-\frac{iK}{2\pi}\frac{H}{P}x^2\right)}
{e^{iKx}-1}
\frac{\prod_{j=1}^{n}\sin\left(\frac{x}{p_j}\right)}
{\sin^{n+2g-2}(x)}.
\nonumber
\end{eqnarray}

The residue is a polynomial $P_{2g-2}(K)$ in $K$ of order $2g-2$:
\begin{eqnarray}
\lefteqn{
Z^{(\cM_0^\prime)}_{l\neq 0}(X_g;k)=
}
\label{4.*10}\\
&&=-\frac{i\pi}{2^{g-1}}
\frac{\sign{P}}{\sqrt{|P|}}
e^{i\pi\frac{3}{4}\sign{\frac{H}{P}}}
\exp\frac{i\pi}{2K}\left(\frac{H}{P}-3\sign{\frac{H}{P}}
-12\sum_{j=1}^{n}s(q_j,p_j)\right)
K^{g-1}P_{2g-2}(K),
\nonumber
\end{eqnarray}
here
\qq
P_{2g-2}(K)=\frac{i^{2g-3}K^{2g-2}}{\pi P}\frac{B_{2g-2}}{(2g-2)!}
-\frac{i^{2g-4}K^{2g-3}H}{2\pi^2 P^2}
\frac{B_{2g-4}}{(2g-4)!}+\ldots,
\label{4.*11}
\qqq
$B_{n}$ are Bernoulli numbers.
\label{p4.2}
\end{proposition}
%***************************
%
The following statement summarizes our calculations:
%
%****************************
\begin{proposition}
The contribution of the connected piece $\cM_0^\prime$ of the moduli
space of flat connections  on a Seifert manifold $\xgf$, which
contains the trivial connection, is given by a sum of two terms
\qq
Z^{(\cM_0^\prime)}(X_g;k)=
Z_{l\neq 0}^{(\cM_0^\prime)}(X_g;k)+
Z_{l=0}^{(\cM_0^\prime)}(X_g;k)
\label{4.15}
\qqq
which are expressed by eqs.~(\ref{4.*09}) and (\ref{4.08}). The first
term is a finite polynomial in $K$, while the second one in an
asymptotic series in $K^{-1}$.
\label{p4.3}
\end{proposition}
%*******************************
%

There is an obvious similarity between these results and the
calculations of \cite{Wi2}. Technically it comes from the similarity
between the sum~(\ref{4.6}) and the formula for a partition function
of the 2d gauge theory. The asymptotic expansion of both expressions
can be calculated with the help of the same technical tricks. The
connected piece $\cM_0^\prime$ of the moduli space of flat connections
on $\xgf$ which contains the trivial connection is isomorphic to the
moduli space of trivial connections on $\Sigma_g$. Therefore the
polynomial $P_{2g-2}(K)$ may also be related to some intersection
numbers on $\cM_0^\prime$. Note however, that the degree of this
polynomial is bigger than that of its counterpart in \cite{Wi2}.

Similar to \cite{Wi2}, the $l=0$ contribution~(\ref{4.9}) should be
related to the singularity of $\cM_0^\prime$ at the trivial
connection. It carries the fractional power of $K$, but it is an
asymptotic series rather than one term as in \cite{Wi2}.

It remains to be determined if either of the second terms in
eqs.~(\ref{4.09}) or (\ref{4.11}) might be related to Casson-Walker
invariant of $X_g$. According to C. Lescop \cite{Le2} this invariant
should be zero for $g\geq 2$. Note that when $g\geq 2$,
$Z_{l\neq 0}^{(\cM_0^\prime)}(X_g;k)$
starts dominating $Z_{l=0}^{(\cM_0^\prime)}(X_g;k)$ in the large $k$
limit.

%************************************
\nsection{Discussion}
%************************************
Even apart from invoking the path integral representation of Witten's
invariant, some arguments of Sections~\ref{*2} and \ref{*3} where not
quite rigorous. A more careful study of boundary conditions for
Witten's invariant of manifolds with boundary and their relation to
the boundary conditions used in \cite{RaSi} to define the analytic
torsion, is needed. The formula~(\ref{3.4}) might also require a more
rigorous proof. We take however some encouragement from the fact that
it contains Walker's surgery formula \cite{Wa} for Casson-Walker
invariant, whose correctness has been checked.

The further study of loop expansion for Witten's invariant of
manifolds with boundaries may be a useful tool in understanding
Vassiliev invariants. According to formulas of Section~\ref{*2},
Vassiliev invariants are rearranged and related to Feynman diagrams on
a knot complement. In this approach the emphasis would be on cubic
vertices rather than on chords of the knot diagrams.

The appearance of Casson-Walker invariant as a 2-loop correction to
the contribution of the trivial connection looks strange. After all,
Casson-Walker invariant is rather a ``number'' of flat $SU(2)$
connections than a local property of Chern-Simons action near the
trivial connection. However this fact has its precedent. The 1-loop
correction to the contribution of the trivial connection (as well as
other connections) to the $U(1)$ Witten's invariant of a \rhs\ $M$ is
proportional to the square root of the Reidemeister-Ray-Singer torsion
of $M$, which is known to be equal to $\ohm$. On the other hand, the
order of homology is equal to the number of flat $U(1)$ connections on
$M$ and may be called a $U(1)$ Casson-Walker invariant. Going further
along this way we can expect to find the surgery formulas for
Casson-Walker invariants of other groups among the higher loop pieces
of eq.~(\ref{3.9}).

\section*{Acknowledgements}

I am thankful to D.~Auckley, S.~Axelrod, D.~Bar-Natan, S.~Garoufalidis,
L.~Jeffrey, L.~Kauffman,
E.~Klassen, H.~Saleur, C.~Taubes, O.~Viro, K.~Walker,
and E.~Witten for discussing the various subjects of this work. I am
especially thankful to D.~Freed, N.~Reshetikhin and A.~Vaintrob for
many consultations and encouragement.

This work was supported in part by the National Science Foundation
under Grants No. PHY-92 09978 and 9009850 and by the R.~A.~Welch
Foundation.

%**********************
%\nnsection{Appendix 1}
\nappendix{}
\label{A*1}
%**********************
%\def\theequation{{A1.\arabic{equation}}}
\def\tk{\cK_{m,n}}
\def\datk{\Delta_A(S^3,\tk;\exp(2\pi ia))}
\def\acrl{a^{({\rm cr})}_l}
\def\tg{\tilde{\g}}

\def\ztk{Z_\a(S^3,\tk;k)}
\def\zbtk{Z^{(\pm\a)}_\a(S^3,\tk;k)}
\def\ta{\tilde{\a}}
\def\ztrtk{Z_\a^{({\rm tr})}(S^3,\tk;k)}
\def\zltk{Z_\a^{(l)}(S^3,\tk;k)}
\def\bstl{\b^{({\rm st})}_l}
\def\bst-l{\b^{({\rm st})}_{-l}}

We illustrate some calculations of Section~\ref{*2} by considering an
example: a large $k$ asymptotics of the Jones polynomial of the type
$(m,n)$ torus knot $\tk$ in $S^3$.
The $1/K$ expansion of this polynomial has been worked out by
H.~Morton in \cite{Mo}.
Here we take a limit in which the
ratio $a=\a/K$ is kept constant as $k\rightarrow\infty$. This allows us to
identify the contributions of various flat connections in the knot complement.
According to the Proposition~\ref{p2.2}, the contribution of the reducible
flat connection will provide the usual $1/K$ expansion of the Jones polynomial.

The Jones polynomial $\ztk$ is expressed as a sum
\begin{eqnarray}
\lefteqn{
\ztk=
}
\label{2.t1}
\\
&&=\frac{1}{2}\sum_{\mu=\pm 1}
\sum_{\stackrel{\b=-\a+1}{\b-\a{\rm\;odd}}}^{\a-1}
\sqrt{\frac{2}{K}}\sin\left[\frac{\pi}{K}(m\b+\mu)\right]
\exp\frac{i\pi}{2K}\left[\frac{n}{m}(m\b+\mu)^2-
\frac{n}{m}-nm(\a^2-1)\right].
\nonumber
\end{eqnarray}
The sum over $\b$ can be converted into an integral with the help of
the Poisson resummation formula. In this particular case
\qq
\sum_{\stackrel{\b=-\a+1}{\b-\a{\rm\;odd}}}^{\a-1}
=\frac{1}{2}\sum_{l\in\ZZ}\int_{-\a}^{\a}d\b\,
\exp[i\pi l(\a+\b+1)],
\label{2.t2}
\qqq
so that
\begin{eqnarray}
\lefteqn{
\ztk=\frac{1}{8i}\sqrt{\frac{2}{K}}\sum_{\mu_{1,2}=\pm 1}\mu_1\mu_2
}
\label{2.t03}
\\
&&\times\sum_{l\in\ZZ}
\int_{-\a}^{\a}d\b\,
\exp\frac{i\pi}{K}\left[\frac{mn}{2}\b^2+
(n\mu_1+m\mu_2+Kl)\b+\mu_1\mu_2-\frac{mn}{2}(\a^2 -1)+Kl(\a+1)\right]
\nonumber
\end{eqnarray}
To find the large $k$ asymtotics of this expression we apply a
stationary phase approximation to the integral over $\b$ in the way
similar to~\cite{Ro1}. We start with the contribution of the boundary
points $\b=\pm\a$. They contribute the same amount due to the
symmetry of the integrand, so we simply double the contribution of
the point $\b=-\a$. After shifting the integration variable
$\b\longrightarrow\b-\a$ and some additional transformations
we get the following formula:
\begin{eqnarray}
\lefteqn{
\zbtk=}
\label{2.t3}\\
&&=\frac{1}{4i}\sqrt{\frac{2}{K}}
\sum_{\mu_{1,2}=\pm 1}\mu_1\mu_2
\exp\frac{i\pi}{K}
\left[-\a(\mu_1 n+\mu_2 m) +
\frac{m^2 n^2 -m^2 -n^2}{2mn}\right]
\sum_{l\in\ZZ}e^{i\pi l}
\sum_{j=0}^{\infty}\frac{1}{j!}
\left(\frac{Kmn}{2\pi i}\right)^j
\nonumber\\
\shift{
\left.\times
\partial_{\e}^{(2j)}\left\{
\exp\left[\frac{i\pi}{K}\e
\left(\frac{\mu_1}{m}+\frac{\mu_2}{n}\right)\right]
\int_{0}^{\infty}d\b\,
\exp\left[\frac{i\pi}{K}\b(\e-mn\a+Kl)\right]\right\}
\right|_{\e=0}
}
\nonumber\\
&&=\frac{1}{4i}\sqrt{\frac{2}{K}}
\sum_{\mu_{1,2}=\pm 1}\mu_1\mu_2
\exp\frac{i\pi}{K}
\left[-\a(\mu_1 n+\mu_2 m) +
\frac{m^2 n^2 -m^2 -n^2}{2mn}\right]
\sum_{j=0}^{\infty}\frac{1}{j!}
\left(\frac{Kmn}{2\pi i}\right)^j
\nonumber\\
\shift{
\left.\times
\partial_{\e}^{(2j)}\left\{
\exp\left[\frac{i\pi}{K}\e
\left(\frac{\mu_1}{m}+\frac{\mu_2}{n}\right)\right]
\left[\frac{iK}{\pi}\sum_{l\in\ZZ}\frac{e^{i\pi l}}
{\e-mn\a-Kl}\right]\right\}\right|_{\e=0}
}
\nonumber\\
&&=
\sqrt{\frac{2}{K}}
\sum_{\mu_{1,2}=\pm 1}\mu_1\mu_2
\exp\frac{i\pi}{K}
\left[
\frac{m^2 n^2 -m^2 -n^2}{2mn}\right]
\sum_{j=0}^{\infty}\frac{1}{j!}
\left(\frac{2\pi i}{K}mn\right)^{-j}
\nonumber\\
\shift{
\times
\partial_{\ta}^{(2j)}\left.\left[
\frac{\sin\left(\frac{\pi}{K}n\ta\right)\,\sin\left(\frac{\pi}{K}m\ta\right)}
{\sin\left(\frac{\pi}{K}mn\ta\right)}\right]\right|_{\ta=\a}
}
\nonumber
\end{eqnarray}
If we substitute eq.~(\ref{2.03}) then we find that in the 1-loop
approximation
\qq
\zbtk=\sqrt{\frac{2}{K}}\frac{\sin(\pi na)\sin(\pi ma)}
{\sin(\pi mna)}+\cO(K^{-\frac{3}{2}}).
\label{2.t4}
\qqq
This is a particular case of relation~(\ref{2.7}) if we recall the
formula for the Alexander polynomial of the torus knot
\qq
\datk=\frac{\sin(\pi mna)\sin(\pi a)}
{\sin(\pi ma)\sin(\pi na)}.
\label{2.t04}
\qqq
Thus we see that the contribution of the boundary points $\b=\pm\a$ is
in fact the contribution of the reducible flat connection in the knot
complement satisfying the boundary condition~(\ref{2.5}). The final
expression of eq.~(\ref{2.t3}) is well defined not only for $\a$ of
order $K$ (where it was derived) but also for $\a$ of order 1  where
it becomes a formula for $\ztrtk$.

The stationary phase points of the integral~(\ref{2.t03})
\qq
\bstl=-\frac{Kl}{mn}
\label{2.t5}
\qqq
also contribute to the Jones polynomial $\ztk$. The contributions of
the points $\bstl$ and $\bst-l$ are equal and should be combined into
\begin{eqnarray}
\lefteqn{\zltk=}
\label{2.t6}\\
&&=\frac{1}{4i}\sqrt{\frac{2}{K}}\sum_{\mu_{1,2}=\pm 1}\mu_1\mu_2
\int_{-\infty}^{+\infty}d\b\,\exp\frac{i\pi}{K}
\left[\frac{mn}{2}\b^2+(\mu_1 n+\mu_2 m+Kl)+\mu_1\mu_2
-\frac{mn}{2}(\a^2-1)
\nonumber\right.\\
\shift{
\left.+Kl(\a+1)\right]
}
\nonumber\\
&&=\frac{i}{2}e^{i\frac{\pi}{4}\sign{mn}+i\pi l}
\exp\left[-i\pi K\frac{mn}{2}\left(a-\frac{l}{mn}
\right)^2\right]\,
\frac{4\sin\left(\pi\frac{l}{m}\right)\sin\left(\pi\frac{l}{n}\right)}
{\sqrt{|mn|}}
\nonumber\\
\shift{
\times
\exp\left[\frac{i\pi}{2K}
\left(\frac{m^2 n^2-m^2-n^2}{mn}\right)\right].
}
\nonumber
\end{eqnarray}

The contributions $\zltk$ come from the irreducible flat connections
in the knot complement satisfying eq.~(\ref{2.5}).
The classical exponent as well as the ingredients of the 1-loop
formula~(\ref{1.14}) can be easily identified in the
expression~(\ref{2.t6}). Note that the whole expression is 2-loop
exact similar to the results of \cite{Ro1}.

A particular value of $l$ may contribute to the integral~(\ref{2.t03})
only if the point $\bstl$ lies within the integration range, that is
\qq
0< l<mna.
\label{2.t7}
\qqq
This condition can not be satisfied for sufficiently small values of
$a$. We see that the first irreducible connection appears only for
$a>\frac{1}{mn}$. New irreducible connections emerge at critical
values
\qq
\acrl=\frac{l}{mn},\;\;\;0<l<mn.
\label{2.t8}
\qqq
These are also the zeros of the Alexander polynomial~(\ref{2.t04}).
This is not surprising (see e.g. \cite{Kl}, \cite{FrKl}), since
the zeros signal the presence of a zero mode
in one of the determinants comprising the Ray-Singer analytic torsion.
This mode is responsible for the ``off-diagonal'' deformation of the
flat reducible connection.

The formula~(\ref{2.t3}) becomes singular near the critical
points~(\ref{2.t8}). However
in this case it only means that the calculation of
the contribution of the boundary points $\b=\pm\a$ has to be modified
when one of the stationary phase points~(\ref{2.t5}) is close to the
boundary. More specifically, the integral over $\b$ in
eq.~(\ref{2.t3}) has to be recalculated for $l=l_0$ if
\qq
\a=K\frac{l_0}{mn}+\g, \;\;\;\;\g\ll\sqrt{K},\;\;0<l_0<mn.
\label{2.t9}
\qqq
As a result, the combined contribution of the stationary phase points
$\b^{({\rm st})}_{\pm\l_0}$ and boundary points $\pm\a$ to $\ztk$ is
given by the formula
\begin{eqnarray}
\lefteqn{Z_{K\frac{l_0}{mn}+\g}^{(l_0,\pm\a)}
(S^3,\tk;k)=}
\label{2.t10}\\
&&=ie^{i\frac{\pi}{4}\sign{mn}+i\pi l_0}
\frac{\sin\left(\pi\frac{l_0}{m}\right)\,\sin\left(\pi\frac{l_0}{n}\right)}
{\sqrt{|mn|}}\exp\left[\frac{i\pi}{2k}\left(-mn\g^2
+\frac{m^2n^2-m^2-n^2}{mn}\right)\right]
\nonumber\\
&&+\sqrt{\frac{2}{k}}e^{i\pi l_0}
\exp\left[\frac{i\pi}{K}\left(\frac{m^2n^2-m^2-n^2}{mn}\right)\right]
\sum_{j=0}^{\infty}\frac{1}{j!}\left(\frac{2\pi i}{K}mn\right)^{-j}
\nonumber\\
&&\times\left.
\partial^{2}_{\tg}\left[
\frac{\sin\pi\left(\frac{l_0}{m}+\frac{n}{K}\tg\right)\,
\sin\pi\left(\frac{l_0}{n}+\frac{m}{K}\tg\right)}
{\sin\left(\frac{\pi}{K}mn\tg\right)}
-\frac{K}{\pi}\frac{\sin\left(\pi\frac{l_0}{m}\right)\,
\sin\left(\pi\frac{l_0}{n}\right)}{mn\tg}\,
\exp\left(-\frac{i\pi}{K}mn\tg\g\right)\right]\right|_{\tg=\g},
\nonumber
\end{eqnarray}
which demonstrates the smooth behavior of $\ztk$ in the vicinity of
critical points~(\ref{2.t8}).

The following proposition summarizes our calculations:
%*********
\begin{proposition}
A large $k$ asymptotics of the Jones polynomial of a
type $(m,n)$ torus knot
in the limit when $a=\frac{\a}{K}$ is kept fixed, contains the
contribution of the reducible connection~(\ref{2.t3}) as well as the
contributions of irreducible connections~(\ref{2.t6}) for the values
of $l$ satisfying the condition~(\ref{2.t7}).

The formula~(\ref{2.t3})
works also in the limit when $\a$ is fixed. It becomes a
contribution of the trivial connection and provides a usual $1/K$
expansion  of the Jones polynomial $\ztk$.

The expression~(\ref{2.t3}) has a singular behavior at zeros of the
Alexander polynomial~(\ref{2.t04}), however the whole
Jones polynomial is smooth (see eq.~(\ref{2.t10})).
\end{proposition}

\end{document}